\documentclass{article}
\usepackage{spconf,amsmath,graphicx,amssymb,url,cite,comment}

\usepackage{tabularx}
\usepackage{xcolor}
\usepackage{booktabs}
\usepackage{arydshln}
\usepackage{hyperref}

\title{Masked Modeling Duo: Learning Representations\\ by Encouraging Both Networks to Model the Input}

\name{Daisuke Niizumi, Daiki Takeuchi, Yasunori Ohishi, Noboru Harada, and Kunio Kashino}

\address{NTT Corporation, Japan
\vspace{-10pt}}

\begin{document}
\ninept
\maketitle
\begin{abstract} % approximately 100 to 150 words
Masked Autoencoders is a simple yet powerful self-supervised learning method. However, it learns representations indirectly by reconstructing masked input patches.
Several methods learn representations directly by predicting representations of masked patches; however, we think using all patches to encode training signal representations is suboptimal.
We propose a new method, Masked Modeling Duo (M2D), that learns representations directly while obtaining training signals using only masked patches.
In the M2D, the online network encodes visible patches and predicts masked patch representations, and the target network, a momentum encoder, encodes masked patches.
To better predict target representations, the online network should model the input well, while the target network should also model it well to agree with online predictions.
Then the learned representations should better model the input.
We validated the M2D by learning general-purpose audio representations, and M2D set new state-of-the-art performance on tasks such as UrbanSound8K, VoxCeleb1, AudioSet20K, GTZAN, and SpeechCommandsV2.
\end{abstract}


%%%%%%%%%%%%%%%%%%%%%%%%%%%%%%%%%%%%%%%%%%%%%%%%%%%%
\section{Introduction}
\label{sec:intro}
%%%%%%%%%%%%%%%%%%%%%%%%%%%%%%%%%%%%%%%%%%%%%%%%%%%%

%NLP分野でのMasked Language Modelingの大きな成功に続き、画像分野ではMasked Image Modeling (MIM)を用いた自己教師あり学習(SSL)の様々な研究が発展し有望な結果を示している。
%中でもMasked Autoencodersは数多くの研究をインスパイアし、画像分野だけでなく音響分野にもその影響を受けた研究が広がっている。
%Following the great success of Masked Language Modeling in the natural language processing domain, various Self-Supervised Learning (SSL) studies with Masked Image Modeling (MIM) have progressed and shown promising results in the image domain.
Recently, self-supervised learning (SSL) methods using masked image modeling (MIM) have progressed and yielded promising results in the image domain.
Among them, Masked Autoencoders\cite{he2022masked} (MAE) have inspired numerous subsequent studies and influenced not only the image domain\cite{tao2022MIM:SIM,assran2022MIM:MSN,chen2022MIM:CAE,elnouby2021MIM:SplitMask} but also the audio domain\cite{huang2022maskedlisten,Baade2022MAE-AST,chong2022maskspec,niizumi2022msm-mae}.
% Removed zhang2022MAEsurvey -> too new, 2208

%MAEは入力を格子状に分割したパッチの大部分をマスクし、残りの少量の可視パッチを使ってマスクパッチを再構成するタスクにより、入力を効果的にモデル化する表現を学習させる。しかしながら、MAEは元データと再構成結果の間で損失を計算するため、表現を学習する目的には最適ではないと考えられる。
An MAE effectively learns a representation by reconstructing a large number (i.e., 75\%) of masked input patches using a small number of visible patches, encouraging the learned representation to model the input. However, it learns representations indirectly by minimizing the loss between the original input and the reconstructed result, which may not be optimal for learning a representation.

\begin{figure}[htbp]
  \centering
  \includegraphics[width=1.0\columnwidth]{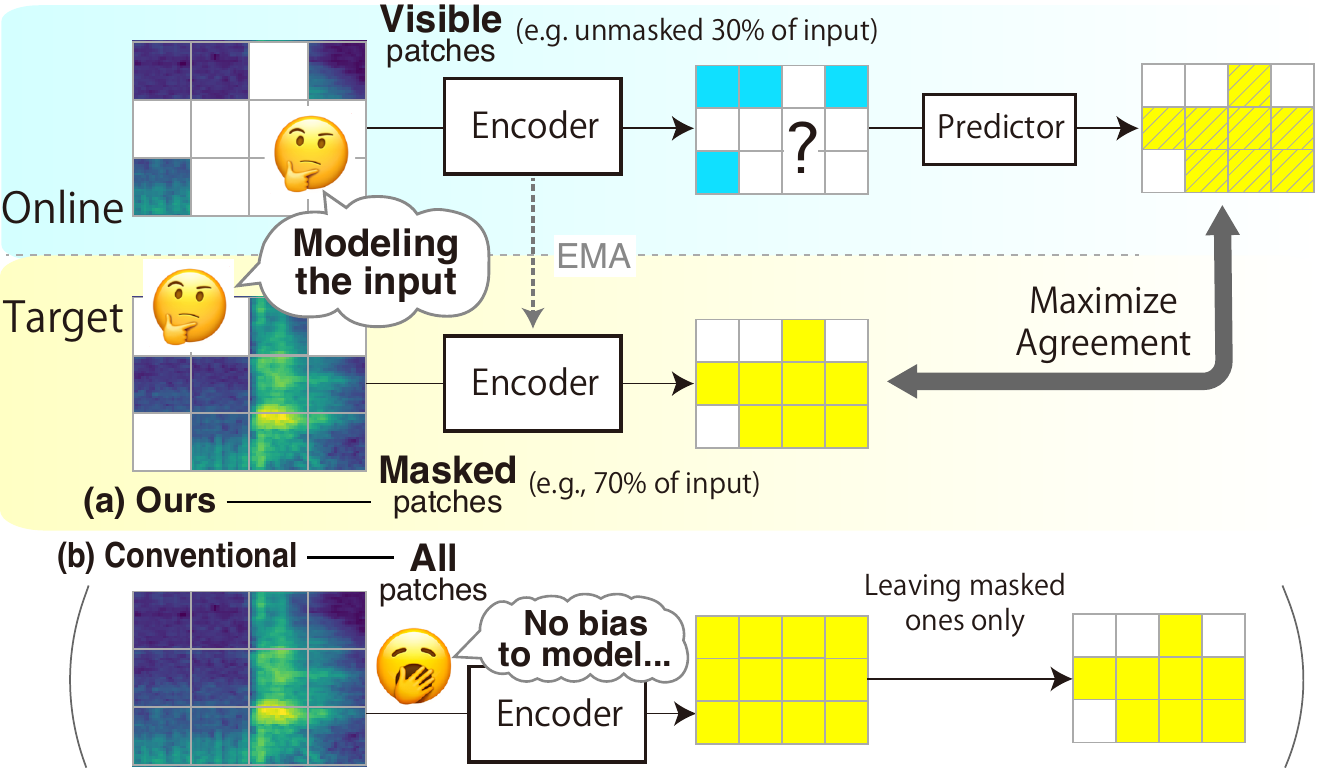} 
  \vspace{-15pt}
  \caption{
  %オンラインは可視パッチを表現にエンコードし、マスクパッチの表現を予測する。ターゲットはマスクパッチを表現にエンコードする。我々のフレームワークはこの２つ潜在表現の一致を最大化することで、入力データの再構成を行わずに表現を学習する。従来手法と異なりターゲットにマスクパッチのみを与えることで、両側から表現に入力をモデル化するように促す。
  M2D pre-training scenario. The online network encodes visible patches and predicts masked patch representations, while the target network encodes masked patches. The M2D maximizes the agreement between these two outputs to learn representations. % without reconstructing the input.
  We provide only the masked patches to the target illustrated as (a), unlike conventional methods (e.g., data2vec\cite{baevski2022data2vec}) depicted as (b), encouraging representations to model the input from both the online and target networks.
  }
  \label{fig:scenario}
  \vspace{-10pt}
\end{figure}

%これに対して、いくつかの既存手法が表現の直接学習を実現しており、典型的にシャムアーキテクチャでモメンタムエンコーダを利用することで教師信号であるマスクパッチの表現を得る。このとき、表現をエンコードするために全ての入力パッチが利用される。
In contrast, several previous methods\cite{tao2022MIM:SIM,assran2022MIM:MSN,baevski2022data2vec} achieve direct learning of representations, typically by using a momentum encoder in a Siamese architecture\cite{Xinlei2021SimSham} to obtain the masked patch representations as a training signal. In this case, all input patches are used to encode these representations, not encouraging to model the input.

%我々は全てのパッチを使うよりマスクパッチだけをエンコードして得る表現のほうが教師信号として有益になりえるという仮説を立てる。
%MAEではエンコーダに少量の可視パッチしか与えないことで入力信号のモデル化を促すのに対して、すべての入力をエンコーダに与えるとモデル化を促すバイアスを享受できないと考えられる。
We hypothesize that the learned representation would become more useful if the training signal were encoded using only masked patches instead of all patches in order to encourage modeling the input in the training signal.
While an MAE effectively encourages modeling the input signal by limiting the number of visible patches fed to the encoder, using all the input patches to obtain a training signal does not benefit from the inductive bias of the MAE.

%本研究では、新たなSSL手法M2Dを提案する。
In this paper, we propose a new method, Masked modeling duo (M2D), that learns representations directly by predicting the representations of masked patches from visible patches only.
%本研究では、シャムアーキテクチャを構成するオンラインとターゲット２つのネットワークで入力のモデル化を促進する新たなSSL手法M2Dを提案する。
%In this paper, we propose a new method, Siamese Masked Autoencoders (M2D), that encourages modeling of the input in the two networks, online and target, forming the Siamese architecture.
%我々の手法は、可視パッチだけからマスクパッチの表現を予測することで表現を直接学習する。そのマスクパッチの表現は、従来手法のような全ての入力パッチではなく、マスクパッチのみからエンコードする。我々の手法はMAEにターゲットモメンタムエンコーダを追加するが、全体のフレームワークは依然シンプルである。
%Our method learns representations directly by predicting the representations of masked patches from visible patches only, as illustrated in Fig. \ref{fig:scenario}.
As illustrated in Fig.\ref{fig:scenario},
the target representations are encoded from only the masked patches, not from all the input patches as they are in the previous methods\cite{tao2022MIM:SIM,assran2022MIM:MSN,baevski2022data2vec}.
Although our method adds a target momentum encoder to the MAE, the entire framework remains simple.

The M2D promotes complementary input modeling by feeding mutually exclusive patches to its networks.
%例えば2つの要素(S1とS2)で構成される心音の入力に対して予測誤差を小さくするためには、オンラインはマスクされたS1の表現を予測するために、与えられるS2のパッチを心音全体の一部としてモデル化した表現にエンコードする必要がある。一方で、ターゲットがエンコードするマスクされたS1の表現は、心音全体の一部としてのエンコードできれば予測と一致しやすいと考えられる。このように我々の手法は両側からモデル化を促進する。
For example, to reduce the prediction error for a heartbeat audio input consisting of two sounds (S1 and S2), the online network should encode the given visible patches around S2 into representations modeled as part of the whole heartbeat in order to predict the representations of masked patches around S1. Conversely, the masked patch representations around S1 encoded by the target network are more likely to agree with the prediction if it is encoded as part of the whole heartbeat. Therefore, our method encourages input modeling from both sides.

%実験では音響信号を入力として汎用音響信号表現の学習を行い、我々の手法を検証する。実験により、表現を直接学習すること、ターゲットにマスクパッチだけを与えることの有効性を確認した。また、複数のタスクでSOTA性能を示した。
In our experiments, we validated our method by learning a general-purpose audio representation using an audio spectrogram as input and confirmed the effectiveness of learning the representation directly and providing only masked patches to the target. In addition, M2D set new state-of-the-art (SOTA) performance on several audio tasks.
Our code is available online\footnote{\url{https://github.com/nttcslab/m2d}}.

%%%%%%%%%%%%%%%%%%%%%%%%%%%%%%%%%%%%%%%%%%%%%%%%%%%%
\section{Related Work}
%%%%%%%%%%%%%%%%%%%%%%%%%%%%%%%%%%%%%%%%%%%%%%%%%%%%
%本研究はMIMとしてMasked Autoencoders (MAE)に、また、ターゲットネットワークを使い潜在表現を直接学習するフレームワークとしてBootstrap Your Own Latent (BYOL) にそれぞれインスパイアされた。MAEは入力データの再構成タスクで学習するのに対して、本研究はマスクされた潜在表現の予測タスクで学習する。BYOLはデータ拡張に不変な表現の学習を行うフレームワークであり、それに対して我々はデータ拡張を利用しない。
This study was inspired by MAE\cite{he2022masked} for an MIM and Bootstrap Your Own Latent\cite{grill2020byol} (BYOL) as a framework for directly learning latent representations using a target network.
An MAE learns to reconstruct the input data, whereas our M2D learns to predict masked latent representations.
BYOL differs from ours in that it is a framework for learning representations invariant to data augmentation.

%ターゲットネットワークを利用する方法
SIM\cite{tao2022MIM:SIM}, MSN\cite{assran2022MIM:MSN}, and data2vec\cite{baevski2022data2vec} learn to predict masked patch representations using a target network, but, unlike ours, all input patches are fed to the target.
CAE\cite{chen2022MIM:CAE} and SplitMask\cite{elnouby2021MIM:SplitMask} encode target representation using only masked patches, which is similar to ours but without the use of a target network.
While SIM, MSN, CAE, and SplitMask learn image representations, data2vec also learns audio representations. %can learn representations regardless of modalities.

%本稿では、汎用音響信号表現の学習により実験を行う。マスク入力を用いて学習する音声音響手法は様々に提案されている。特に、XXはMAEを適用して音響信号の表現を学習する。
In this work, we experimented through learning general-purpose audio representations.
To learn speech and audio, various methods learn representations using masked input, such as Mockingjay\cite{Liu2020Mockingjay}, wav2vec2\cite{baevski2020wav2vec2}, HuBERT\cite{Hsu2021HuBERT}, and BigSSL\cite{zhang2021bigssl} for speech, and SSAST\cite{gong2022ssast} for audio.
Methods more closely related to ours are MAE-AST\cite{Baade2022MAE-AST}, MaskSpec\cite{chong2022maskspec}, MSM-MAE\cite{niizumi2022msm-mae}, and Audio-MAE\cite{huang2022maskedlisten}, which adapt MAE to learn audio representations. However, they differ from our method in that they do not use a target network.

%音を学習するその他のSSL手法にはAAとBBが挙げられ、特にCC、DD、EEはBYOLを学習フレームワークとして学習する。
Other SSL methods for learning audio representations include Wang et al.\cite{wang2022universal} and DeLoRes\cite{Ghosh2022DeLoRes}, and especially BYOL-A\cite{niizumi2022byol-a}, BYOL-S\cite{scheidwasserclow2021serab}, and ATST\cite{Li2022ATST}, which use BYOL as the learning framework; they do not mask the input.
%教師あり学習手法としては、AST、EAT、PaSST、HTS-ATがSOTA性能を示している。
For supervised learning, AST\cite{gong2021ast}, EAT\cite{gazneli2022EAT}, PaSST\cite{Koutini2022passt}, and HTS-AT\cite{Chen2022HTS-AT} have shown SOTA performance.

%%%%%%%%%%%%%%%%%%%%%%%%%%%%%%%%%%%%%%%%%%%%%%%%%%%%
\section{Masked Modeling Duo}
\begin{figure}[tbhp]
  \vspace{-5pt}
  \centering
  \includegraphics[width=0.88\columnwidth]{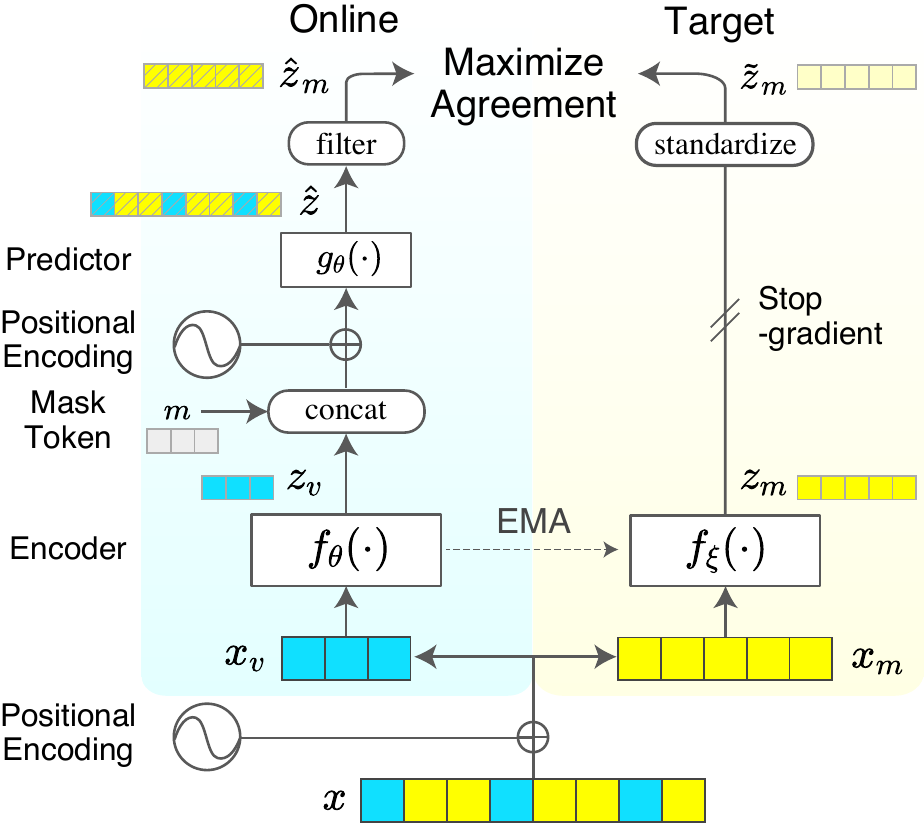}
  \vspace{-3pt}
  \caption{Overview of the M2D framework.}
  \label{fig:system}
  \vspace{-5pt}
\end{figure}

%%%%%%%%%%%%%%%%%%%%%%%%%%%%%%%%%%%%%%%%%%%%%%%%%%%%
%我々の手法は可視パッチだけを使ってマスクパッチの表現を予測することで表現を学習する。図Xに示されるように、フレームワークはオンライン、ターゲット２つのネットワークで構成される。
Our method learns representations by using only visible patches to predict the masked patch representations. As shown in Fig. \ref{fig:system}, it consists of two networks, referred to as the online and target networks.

\vspace{0.1cm}
\noindent\textbf{Processing input}\hspace{0.2cm}
%まず、フレームワークは入力データX(オーディオスペクトログラムや画像など)を格子状のパッチに分割、位置エンコーディングを付与し、そのうちマスク率に応じた個数をランダムに選んでマスクパッチXMとし、残りを可視パッチXVとして割り当てる。
The framework partitions the input data $x$ (audio spectrogram, image, etc.) into a grid of patches, adds positional encoding, and randomly selects a number of patches according to a masking ratio as masked patches $x_m$ (e.g., 60\% of the input) and the rest as visible patches $x_v$ (e.g., the remaining 40\%).
%Random sampling is structure-agnostic, as in MAE\cite{he2022masked}.
While we use the same positional encoding as MAE\cite{he2022masked}, we tested various masking ratios as discussed in Section \ref{exp:mask-ratio}.

%\vspace{0.1cm}
\noindent\textbf{Online and target networks}\hspace{0.2cm}
%オンラインネットワークは、パラメーター集合thetaで定義され、エンコーダFSを用いて可視パッチXVを潜在表現ZVにエンコードする。
The online network, defined by a set of weights $\theta$, encodes the visible patches $x_v$ using the online encoder $f_\theta$ into the representation $z_v = f_\theta(x_v)$.
%ZVにマスクパッチMを追加、位置エンコーディング$p$を足し合わせ、予測器GSを使って入力全体の潜在表現ZZを予測する。
It concatenates shared, learnable masked tokens $m$ to $z_v$, adds the position encoding $p$, and predicts  $\hat{z}$, representations of entire input patches, using the predictor $g_\theta$.
\begin{equation}
  \hat{z} = g_\theta(\text{concat}(z_v, m) + p)
\end{equation}
%その後、予測結果をフィルタリングしてマスクパッチに対応する表現のみをZZMとして出力する。
It then filters the prediction result $\hat{z}$ to output $\hat{z}_m = \{\, \hat{z}[i] \mid i \in I_M \,\}$, containing only masked patch representations, where $I_M$ is the set of indices of the masked patches.
%It then outputs the representation $\hat{z}_m = \{\, \hat{z}[i] \mid i \in V \,\}$, corresponding to the masked patches among the prediction results $\hat{z}$, where $V$ is the set of indices of the masked patches.
%\hat{z}_m = \{\hat{z}[i]\}, i \in V,

%ターゲットネットワークはパラメータXIで定義され、パラメータ以外はオンラインエンコーダと同一のエンコーダFXのみを持つ。ネットワークはFXを用いてマスクパッチXMをエンコードした表現ZMを出力する。
The target network is defined by parameter $\xi$ and consists only of momentum encoder $f_\xi$, which is identical to the online encoder except for the parameter.
The network encodes masked patches $x_m$ using $f_\xi$ to output the representation $z_m = f_\xi(x_m)$.
%我々はMAEと同様にZMを標準化する。ターゲットの標準化することで、学習が安定することを実験的に確認した。
We then standardize $z_m$ to $\tilde{z}_m = ({z_m - \text{mean}{(z_m)}})/{\sqrt{\text{var}{(z_m)}}}$, for stabilizing the training, which we empirically confirmed in preliminary experiments, rather than for performance gain as in MAE.
%We then calculate $\tilde{z}_m$, a standardized $z_m$, for stabilizing training which we empirically confirmed in the preliminary experiments, rather than for performance gain as in MAE.
%\begin{equation}
%  \tilde{z}_m = ({z_m - \text{mean}{(z_m)}})/{\sqrt{\text{var}{(z_m)}}},
%\end{equation}

\vspace{0.1cm}
\noindent\textbf{Calculating loss}\hspace{0.2cm}
%損失はオンラインの予測出力ZZMに対してターゲットの出力ZMを教師信号として計算される。
The loss is calculated using the standardized target output $\tilde{z}_m$ as a training signal against the online prediction output $\hat{z}_m$.
%BYOLに則り、L2ノーマライズしたZZMとHZMのmean square error (MSE)により損失Lを計算する。
Inspired by BYOL\cite{grill2020byol}, we calculate the loss $L$ by the mean square error (MSE) of $l_2$-normalized $\hat{z}_m$ and $\tilde{z}_m$.
\begin{equation}
L \triangleq ||l_2(\hat{z}_m) - l_2(\tilde{z}_m)||^2_2 = 2 - 2 \cdot \frac{\langle \hat{z}_m, \tilde{z}_m \rangle }{||\hat{z}_m||_2 \cdot ||\tilde{z}_m||_2},
\label{eq:eq-byol-mse}
\end{equation}
where $\langle\cdot, \cdot\rangle$ denotes the inner product.

\vspace{0.1cm}
\noindent\textbf{Updating network parameters}\hspace{0.2cm}
Our framework updates parameters $\theta$ and $\xi$ after each training step. It updates $\theta$ only by minimizing the loss $L$ as depicted by the stop-gradient in Fig. \ref{fig:system}, whereas updating $\xi$ is based on a slowly moving exponential average of $\theta$ with a decay rate $\tau$:
\begin{equation}
    \xi \leftarrow \tau \xi + (1 - \tau) \theta
\end{equation}

%オンラインネットワークのゆっくりとした移動平均をbootstrapすることで、不十分な解に崩壊することを避け、有益な表現の学習を導けることが経験的に示されている。
It has been empirically shown that stop-gradient operation can avoid collapsing to an uninformative solution, and the moving-average behavior may lead to learning effective representations\cite{Xinlei2021SimSham}.
%提案手法は、モメンタムターゲットでエンコードする表現とオンラインでエンコードする表現の表現の一致が最大化するよう表現を学習する。
%Our method learns representations by maximizing the agreement between the representations encoded at the momentum target and those encoded at the online networks.
After the training, we transfer only the $f_\theta$ as a pre-trained model.

%It has been empirically shown that the combination of adding the predictor to the online network and using the moving average of the online network parameters as the target network encourages encoding more and more information within the online projection and avoids collapsed representations such as constant value\cite{grill2020byol}.

%%%%%%%%%%%%%%%%%%%%%%%%%%%%%%%%%%%%%%%%%%%%%%%%%%%%
\section{Experiments} \label{sec:experiments}
%%%%%%%%%%%%%%%%%%%%%%%%%%%%%%%%%%%%%%%%%%%%%%%%%%%%
%我々の提案を段階を経た実験により有効性を検証した。XX章で我々の提案とMAEとの比較により表現を直接学習することの有効性を、XX章でマスクパッチのみをターゲットに与えることの有効性を検証した。XX章では様々なマスク率で評価し、XX章ではSOTAとの比較を行う。
We validated the M2D step-by-step experiments that examined the effectiveness of learning representations directly by comparing our M2D with MAE (Section \ref{exp:compare-w-mae}), the effectiveness of feeding only masked patches to the target (Section \ref{exp:target-input}), the impact of various masking ratios (Section \ref{exp:mask-ratio}), and comparing ours with SOTA (Section \ref{exp:sota}).

%全ての実験において、我々の手法を音声スペクトログラムを入力とするMasked Spectrogram Modeling (MSM)に応用し、汎用的な音声表現を学習する。
In all experiments, we applied our M2D to masked spectrogram modeling (MSM)\cite{niizumi2022msm-mae}, with an audio spectrogram as input to learn general-purpose audio representations.
%環境音、音声、音楽の下流タスクを用いて、Linear evaluation、fine-tuning両方で事前学習後のモデルの性能を評価した。
We evaluated the performance of pre-trained models in both a linear evaluation and fine-tuning on a variety of audio downstream tasks spanning environmental sounds, speech, and music.

\begin{table*}[htb!]
\vspace{-20pt}
\caption{Fine-tuning and linear evaluation results. AS20K result is an mAP, and all others are accuracies (\%) with 95\% CI.}
\label{tab:results-w-mae}
\centering
\resizebox{\textwidth}{!}{%
\begin{tabular}{lllllllllllllll} \toprule
 & \multicolumn{4}{c}{(a) Fine-tuning} & \multicolumn{10}{c}{(b) Linear evaluation}\\
 \cmidrule(lr){2-5} \cmidrule(lr){6-15} 
 &&&& &  \multicolumn{2}{c}{Env. sound tasks} & \multicolumn{4}{c}{Speech tasks} & \multicolumn{3}{c}{Music tasks} \\
 \cmidrule(lr){6-7} \cmidrule(lr){8-11} \cmidrule(lr){12-14} 
Model & AS20K & ESC-50 & SPCV2 & VC1 &   ESC-50 &    US8K &    SPCV2 &    VC1 &     VF &    CRM-D &    GTZAN &     NSynth &      Surge & Avg.\\
\midrule

MSM-MAE\cite{niizumi2022msm-mae} &  36.7 {\fontsize{6pt}{6pt}\selectfont $\pm$ 0.5} &  94.0 {\fontsize{6pt}{6pt}\selectfont $\pm$ 0.2} &  98.4 {\fontsize{6pt}{6pt}\selectfont $\pm$ 0.1} &\textbf{95.3 {\fontsize{6pt}{6pt}\selectfont $\pm$ 0.1}}&  88.6 {\fontsize{6pt}{6pt}\selectfont $\pm$ 1.5} &  86.3 {\fontsize{6pt}{6pt}\selectfont $\pm$ 0.3} &  94.5 {\fontsize{6pt}{6pt}\selectfont $\pm$ 0.2} &  72.2 {\fontsize{6pt}{6pt}\selectfont $\pm$ 0.2} &  97.5 {\fontsize{6pt}{6pt}\selectfont $\pm$ 0.1} &  70.2 {\fontsize{6pt}{6pt}\selectfont $\pm$ 1.1} &  78.4 {\fontsize{6pt}{6pt}\selectfont $\pm$ 2.8} &  75.9 {\fontsize{6pt}{6pt}\selectfont $\pm$ 0.2} &\textbf{42.5 {\fontsize{6pt}{6pt}\selectfont $\pm$ 0.7}}&  78.5 {\fontsize{6pt}{6pt}\selectfont $\pm$ 0.8} \\
\addlinespace[-0.02cm] \hdashline \addlinespace[0.05cm]

M2D \small{ratio=0.6} &  36.8 {\fontsize{6pt}{6pt}\selectfont $\pm$ 0.1} &  94.7 {\fontsize{6pt}{6pt}\selectfont $\pm$ 0.3} &\textbf{98.5 {\fontsize{6pt}{6pt}\selectfont $\pm$ 0.0}}&  94.8 {\fontsize{6pt}{6pt}\selectfont $\pm$ 0.1} &  89.7 {\fontsize{6pt}{6pt}\selectfont $\pm$ 0.2} &\textbf{87.6 {\fontsize{6pt}{6pt}\selectfont $\pm$ 0.2}}&\textbf{95.4 {\fontsize{6pt}{6pt}\selectfont $\pm$ 0.1}}&\textbf{73.1 {\fontsize{6pt}{6pt}\selectfont $\pm$ 0.1}}&\textbf{97.9 {\fontsize{6pt}{6pt}\selectfont $\pm$ 0.1}}&\textbf{71.7 {\fontsize{6pt}{6pt}\selectfont $\pm$ 0.3}}&  83.3 {\fontsize{6pt}{6pt}\selectfont $\pm$ 1.0} &  75.3 {\fontsize{6pt}{6pt}\selectfont $\pm$ 0.1} &  41.0 {\fontsize{6pt}{6pt}\selectfont $\pm$ 0.2} &\textbf{79.5 {\fontsize{6pt}{6pt}\selectfont $\pm$ 0.3}}\\
M2D \small{ratio=0.7} &\textbf{37.4 {\fontsize{6pt}{6pt}\selectfont $\pm$ 0.1}}&\textbf{95.0 {\fontsize{6pt}{6pt}\selectfont $\pm$ 0.2}}&  98.5 {\fontsize{6pt}{6pt}\selectfont $\pm$ 0.1} &  94.4 {\fontsize{6pt}{6pt}\selectfont $\pm$ 1.3} &\textbf{89.8 {\fontsize{6pt}{6pt}\selectfont $\pm$ 0.3}}&  87.1 {\fontsize{6pt}{6pt}\selectfont $\pm$ 0.3} &  94.5 {\fontsize{6pt}{6pt}\selectfont $\pm$ 0.1} &  71.3 {\fontsize{6pt}{6pt}\selectfont $\pm$ 0.4} &  97.7 {\fontsize{6pt}{6pt}\selectfont $\pm$ 0.1} &  71.6 {\fontsize{6pt}{6pt}\selectfont $\pm$ 0.3} &\textbf{83.9 {\fontsize{6pt}{6pt}\selectfont $\pm$ 1.4}}&\textbf{76.9 {\fontsize{6pt}{6pt}\selectfont $\pm$ 1.3}}&  41.8 {\fontsize{6pt}{6pt}\selectfont $\pm$ 0.4} &  79.4 {\fontsize{6pt}{6pt}\selectfont $\pm$ 0.5} \\

\bottomrule
\addlinespace[0.05cm]
\end{tabular}
}
\vspace{-15pt}
\end{table*}

%\begin{table}[htb!]
%\caption{Fine-tuning comparison between methods directly related to M2D, pre-trained with a 2s input audio (\%).}
%\label{tab:ft-results-w-mae}
%\centering
%\resizebox{0.8\columnwidth}{!}{%
%\begin{tabular}{lllll}
%\toprule
%\bottomrule\\
%\end{tabular}
%}
%\vspace{-5pt}
%\end{table}

\subsection{Experimental Setup}
%我々はMAEとの比較に重点を置き、可能な限りMAEの実装や設定をそのまま流用した。我々はMAE学習フレームワークにターゲットネットワークを追加実装し、MAEのデコーダをそのまま我々のPredictorとして流用した。我々はすべての実験でエンコーダにはバニラViT-Baseを利用し、パッチサイズを16x16に固定した。マスク率には予備実験で良い性能を示した0.6, 0.7を利用した。
We mainly focused on comparing M2D with an MAE, then adapted MAE implementations and settings with as few changes as possible. We implemented an additional target network on top of the MAE code and adopted the MAE decoder as our predictor $g_\theta$ without changes.
We used vanilla ViT-Base\cite{ViT} with a 768-d output feature as our encoders ($f_\theta$ and $f_\xi$) and fixed the patch size to $16\times 16$ for all experiments.
We tested with masking ratios of 0.6 and 0.7, which showed good performance in preliminary experiments.

%比較するMAEには、4層、6ヘッド、384次元の埋め込みによる小さなデコーダーにすることで音響信号に最適化したMSM-MAEを利用した。予備実験により標準的なMAEの性能を改善できることを確認している。
We used the MSM-MAE\cite{niizumi2022msm-mae} as an MAE for comparison, an MAE variant optimized for MSM by making the decoder smaller with four layers, six heads, and a width (embedding dimension) of 384-d. Other parameters, including a default masking ratio of 0.75, were the same as in the original MAE.
Preliminary experiments verified that the MSM-MAE outperforms a vanilla MAE with an eight-layer decoder.

%MSM-MAEの特徴量計算を採用し、M2Dの特徴量計算をMSM-MAE同様MSMに最適化した。
We employed the MSM-MAE feature calculation in evaluations to optimize M2D representations for MSM as in MSM-MAE.
%MSM-MAEでは、ViTの出力するZBFTDを単純に平均する代わりに、時間フレームごとに各周波数便の埋め込みをつなげることで、ZBTFDを出力する。我々はこの出力を時間方向に平均をとったZZを利用する。
The MSM-MAE outputs $z' \in R^{B \times N_T \times N_F D}$ by concatenating the features of frequency bins for each time frame, instead of simply averaging the $z \in R^{B \times N_F N_T \times D}$ output from ViT, where $B$ is batch size, $N_F$ is the number of patches along frequency, $N_T$ is the number of patches along time, and $D$ is a patch feature dimension. Then, we summarized an audio sample level feature $z'' = 1/N_T\sum_{t=1}^{N_T} z'[t]$, averaging $z'$ over time. The $z''$ becomes a 3,840-d feature, where $D=768$ and $N_F=5$.

%オーディオ処理
We preprocessed audio samples to a log-scaled mel spectrogram with a sampling frequency of 16,000 Hz, window size of 25 ms, hop size of 10 ms, and mel-spaced frequency bins $F=80$ in the range of 50 to 8,000 Hz and normalized them with a dataset statistics.
All downstream task audios were cropped to the dataset's average duration or added with zero padding at the end.
%モデルの入力長より長い音声は重複なしに分割し、それぞれをエンコードしてZZを得たあと時間に沿って平均を取りクリップ表現ZZZを得た。
Long audio samples were split into the input length of the model without overlapping, encoded to $z'$ each, concatenated along time, and averaged over time to an audio sample level feature $z''$.

\vspace{0.1cm}
\noindent\textbf{Pre-training details}\hspace{0.2cm}
%我々は入力信号の時間長が異なる二種類のモデルをそれぞれ学習した。一つはSOTA手法との比較のためATSTと同じ入力信号長である6秒を、もう一つは様々な実験を容易にするため2秒の信号長とし、それぞれをMABL 6s、2sと呼ぶ。
%We trained two models with different input audio durations. One model accepts 6s audio, the same input duration as ATST\cite{Li2022ATST} for comparison with SOTA methods. The other accepts a 2s audio to facilitate various experiments. We denote them as \textit{MABL 6s} and \textit{MABL 2s}, respectively.
%それぞれの入力スペクトログラムのサイズ(Freq. bins x Time frames)は2s、6sそれぞれが80x208、80x608となる。
%The sizes of the input spectrogram (Freq. bins $\times$ Time frames) are $80 \times 608$ and $80 \times 208$ for 6s and 2s models, respectively.
%実験に使用したViTの入力信号の時間長は、SOTA手法との比較のためATSTと同じ6秒とした。入力スペクトログラムのサイズは80x608 (Freq. bins x Time frames)となる。
The input audio duration of the ViT used in the experiments was set to 6s, the same as ATST, for comparison with the SOTA methods. The input spectrogram has a size of $80 \times 608$ (Freq. bins $\times$ Time frames), making $N_F=80/16=5$ and $N_T=608/16=38$ with a patch size of $16\times 16$.

%我々はエポック数300、warm upエポック数20、バッチサイズ2048、base learning rate 3e-4とした他はlearning rate schedulingや最適化を含めすべてMAEと同じ設定とした。ターゲットネットワーク更新のEMAパラメータTAUは学習開始時0.99995から終了時0.99999まで線形補間した。
We set the number of epochs as 300, warm-up epochs as 20, batch size as 2048, and the base learning rate as 3e-4. All other settings were the same as in the MAE, including the learning rate scheduling and optimizer. The EMA decay rate $\tau$ for the target network update was linearly interpolated from 0.99995 at the start of training to 0.99999 at the end.
We used AudioSet\cite{gemmeke2017audioset} as a pre-training dataset with 2,005,132 samples (5,569 h) of 10s audio from the balanced and unbalanced train segments.
We randomly cropped 6s audio from a 10s sample.
%All these settings were commonly used for pre-training M2D and MSM-MAE, except that the EMA decay rate was for M2D only.
All these settings were common in the M2D and MSM-MAE pre-training, except for the EMA decay rate.

\vspace{0.1cm}
\noindent\textbf{Linear evaluation details}\hspace{0.2cm}
%この評価は、事前学習済みモデルの性能を、線形モデルを多様な下流タスクで使い評価する。その際事前学習済みの重みをフリーズし、その上に線形層の分類器を学習する。
In the linear evaluation, we trained a \textit{linear classifier} on top of \textit{frozen} pre-trained models, and tested the performance on a variety of downstream tasks.

%下流タスクや評価の詳細は全て先行研究BYOL-A2022に準ずる。
All evaluation details and downstream tasks are the same as in our previous study\cite{niizumi2022byol-a}. Tasks include environmental sound classification ESC-50\cite{piczak2015esc50} and UrbanSound8K\cite{salamon2014urbansound} (US8K), speech-command classification Speech Commands V2\cite{speechcommandsv2} (SPCV2), speaker identification VoxCeleb1\cite{voxceleb} (VC1), language identification VoxForge\cite{voxforge} (VF), speech emotion recognition CREMA-D\cite{cao2014cremad} (CRM-D), music genre recognition GTZAN\cite{gt2013gtzan}, musical instrument classification NSynth\cite{nsynth2017}, and a pitch audio classification Pitch Audio Dataset (Surge synthesizer)\cite{turian2021torchsynth}.
All the tasks are classification problems, and all the results are accuracies.

\vspace{0.1cm}
\noindent\textbf{Fine-tuning details}\hspace{0.2cm}
%我々は先行研究で共通して使われるタスクを用いて性能を評価する。構成するタスクはLinear evaluationと同じESC-50, SPCV2, VC1に加えて、AudioSetのbalanced train segmentsだけを学習し527マルチラベル分類のmean average pooling (mAP)を結果とするAS20Kである。
We used the tasks commonly used in previous studies: ESC-50, SPCV2, and VC1, the same as in the linear evaluation, plus AudioSet20K (AS20K), which learns only the balanced train segments of AudioSet\cite{gemmeke2017audioset} and results in a mean average pooling (mAP) of multi-label classification with 527 classes.

%Fine-tuningパイプラインはATSTに則る。線形分類器を事前学習済みモデルの上に追加し、全体を学習する。全ての評価は200エポック学習し、learning rateはタスクごとに最適化し、5エポックのwarm upを経たCosine annealingでスケジューリングした。我々はオプティマイザにはSGD、データ拡張にはMixupとRRC、更にPaSSTで提案されたStructured Patchoutを利用した。表Xに実験で使用したパラメータを一覧する。
The fine-tuning pipeline follows ATST. A linear classifier was added on top of the pre-trained model to train the entire network. All evaluations were trained for 200 epochs, and the learning rate was optimized for each task and scheduled with cosine annealing\cite{loshchilov2016sgdr} after five epochs of warm-up. We used SGD and AdamW for the optimizer, Mixup\cite{zhang2018mixup,niizumi2022byol-a}, Random Resize Crop (RRC)\cite{niizumi2022byol-a} for data augmentation, and Structured Patchout (SPO) proposed in PaSST\cite{Koutini2022passt} that masks patches during training. When using the SPO, we used 768-d features calculated by averaging ViT outputs over time because the MSM-MAE feature calculation is not applicable to the masked patches.
Table \ref{tab:ft-parameters} summarizes the settings.

\begin{table}[htb!]
\vspace{-15pt}
\caption{Fine-tuning settings}
\label{tab:ft-parameters}
\centering
\resizebox{0.6\columnwidth}{!}{%
\begin{tabular}{lllll}
\toprule
Parameter & AS20K & ESC-50 & SPCV2 & VC1 \\
\midrule
LR        & 1.0 & 0.5 & 0.5 & 0.001 \\
Optimizer & SGD & SGD & SGD & AdamW \\
Mixup     & 0.3 & 0.0 & 0.3 & 0.0 \\
RRC       & \checkmark & \checkmark & \checkmark & - \\
SPO ratio & 0.5 & 0.5 & 0.5 & 0.0 \\
\bottomrule\\
\end{tabular}
}
\vspace{-23pt}
\end{table}

%%%%%%%%%%%%%%%%%
%\subsection{Experiment: Comparing with MAE} \label{exp:compare-w-mae}
\subsection{Validation of Learning Representations Directly} \label{exp:compare-w-mae}
%%%%%%%%%%%%%%%%%
We validated the effectiveness of learning representations directly instead of through the reconstruction task by comparing the M2D with MAE. Note that the pre-trained models shared exactly the same ViT, enabling us to compare the difference of pre-training schemes.

%MAEとの比較結果を表X示す。(a) Fine-tuning, (b) Linear evaluationどちらの結果も、M2Dが従来のMAEの性能を殆どのタスクで改善することを示している。
Table \ref{tab:results-w-mae} shows the comparison results with the MAE. Both (a) fine-tuning and (b) linear evaluation results show that the M2D improves the performance of the conventional MAE in most tasks.
%一方で、linear evaluationでは音高分類Surgeで、話者分類VoxCeleb1ではfine-tuningで性能が劣化している。これらのタスクでは音高情報が重要と考えられ、逆にそれら以外のタスクでは音高が重要ではない分類タスクであると考えられる。このことはMAEに比べてM2Dの表現の音高に関する感度が鈍くなり、従ってSurge, VoxCeleb1での性能が劣化する一方、逆に音高の違いによらない識別を行うその他のタスクで性能が向上していることを示唆すると考えられる。
However, the performance degrades on Surge (pitch classification) in the linear evaluation and on VoxCeleb1 (speaker classification) in fine-tuning.
For these tasks, pitch information is considered important, while for the other tasks, pitch is considered less important.
This suggests that the M2D representations are less sensitive to pitch than the MAE ones, and thus performance is degraded in Surge and VoxCeleb1, while it is improved in the other tasks that typically discriminate events regardless of pitch.

%これらの結果は、MAEのようにデータの再構成による学習が細部の局所的表現を促進し、潜在表現を直接学習することで抽象的な表現を促進している可能性がある。
These results also suggest that learning by reconstructing data, as in the MAE, may facilitate detailed and local representations, while learning latent representations directly may facilitate more abstract representations.
%In summary, we think that the experimental results validated the effectiveness of the M2D.

%%%%%%%%%%%%%%%%%
%\subsection{Experiment: Input to the target} \label{exp:target-input}
\subsection{Validation of Input to the Target} \label{exp:target-input}
%%%%%%%%%%%%%%%%%
We validate the effectiveness of feeding masked patches only to the target by comparing our proposal with providing all patches to the target, which is employed in methods such as data2vec\cite{baevski2022data2vec}. We compare the average results of the linear evaluation.
%表XXに示されるように、data2vecなどで採用されている全てのパッチを与えるよりも、マスクパッチだけをターゲットに与えるほうが良い結果となり、我々の仮説を裏付けた。
The results for both masking ratios in Table \ref{tab:targ-input-results} shows that giving the target only the masked patch improves the result more than giving it all the patches, confirming our hypothesis.

\begin{table}[htb!]
\vspace{-12pt}
%\caption{Target input ablations: linear evaluation average results (\%).}
\caption{Linear evaluation average results (\%) for target inputs.}
\label{tab:targ-input-results}
\centering
\resizebox{0.85\columnwidth}{!}{%
\begin{tabular}{lcc|cc}
\toprule
 & \multicolumn{2}{c|}{{Input ratio}} & \multicolumn{2}{c}{Masking ratio ($r$)} \\
\cmidrule(lr){2-3} \cmidrule(lr){4-5}
Target input & {Online} & {Target} & 0.6 & 0.7 \\
\midrule
Masked patches only (ours) & $1.0-r$ & $r$ & \textbf{79.5 {\fontsize{6pt}{6pt}\selectfont $\pm$ 0.3}} &  \textbf{79.4 {\fontsize{6pt}{6pt}\selectfont $\pm$ 0.5}} \\
All patches (conventional) & $1.0-r$ & $1.0$ & 79.1 {\fontsize{6pt}{6pt}\selectfont $\pm$ 0.3} & 79.3 {\fontsize{6pt}{6pt}\selectfont $\pm$ 0.3} \\
\bottomrule
\end{tabular}
}
\vspace{-15pt}
\end{table}

%音声タスクでは全てのパッチを与えるほうが良い結果となり、本手法の有効性がタスクによって異なりえることを示している。
%まとめとして、平均的に良い結果を示すことから、一般には本手法が有効であると考える。
%For the speech tasks, feeding all patches was better, indicating that the effectiveness of our choice to provide masked patches only can depend on the task.
%In general, our method is considered effective, given that it showed better results on average.

%\begin{figure}[htbp]
%  \centering
%  \includegraphics[width=0.95\columnwidth]{./figure3targinput.pdf} 
%  \caption{Target input ablations: Linear evaluation results (\%).}
%  \label{fig:targ-input-results}
%  %\vspace{-10pt}
%\end{figure}

\begin{table*}[htb!]
\vspace{-10pt}
\caption{Linear evaluation comparison with SOTA models (\%). Supervised learning methods and non-standard linear evaluation results are grayed out as a reference.}
\label{tab:results-sota-le}
\centering
\resizebox{0.8\textwidth}{!}{%
\begin{tabular}{llllllllll} \toprule
 &  \multicolumn{2}{c}{Env. sound tasks} & \multicolumn{4}{c}{Speech tasks} & \multicolumn{3}{c}{Music tasks} \\
\cmidrule(lr){2-3} \cmidrule(lr){4-7} \cmidrule(lr){8-10} 
Model &    ESC-50 &    US8K &    SPCV2 &    VC1 &     VF &    CRM-D &    GTZAN &     NSynth &      Surge \\
\midrule
Wav2Vec2\cite{baevski2020wav2vec2}$^{\dagger}$  &  \underline{57.6} {\fontsize{6pt}{6pt}\selectfont $\pm$ 0.8} &  \underline{66.9} {\fontsize{6pt}{6pt}\selectfont $\pm$ 0.4} & \textbf{\underline{96.6}} {\fontsize{6pt}{6pt}\selectfont $\pm$ 0.0}&  \underline{40.9} {\fontsize{6pt}{6pt}\selectfont $\pm$ 0.6} & \textbf{\underline{99.2}} {\fontsize{6pt}{6pt}\selectfont $\pm$ 0.1}& \underline{65.5} {\fontsize{6pt}{6pt}\selectfont $\pm$ 1.7}&  \underline{57.8} {\fontsize{6pt}{6pt}\selectfont $\pm$ 1.3} &  \underline{56.6} {\fontsize{6pt}{6pt}\selectfont $\pm$ 0.6} &  \underline{15.2} {\fontsize{6pt}{6pt}\selectfont $\pm$ 0.9} \\
DeLoRes-M\cite{Ghosh2022DeLoRes} & - & 82.7 & 89.7 & 45.3 & 88.0 & - & - & 75.0 & - \\
SF NFNet-F0\cite{wang2022universal} & 91.1 & - & 93.0 & 64.9 & 90.4 & - & - & \textbf{78.2} &- \\
%OpenL3\cite{cramer2019openl3} $^{\dagger}$  & 79.8 & 79.3 & \underline{87.9} & \underline{40.7} & \underline{90.1} & \underline{60.4} & \underline{73.3} & \underline{75.6} & \underline{36.4} \\
BYOL-A\cite{niizumi2022byol-a}      &  83.2 {\fontsize{6pt}{6pt}\selectfont $\pm$ 0.6} &  79.7 {\fontsize{6pt}{6pt}\selectfont $\pm$ 0.5} &  93.1 {\fontsize{6pt}{6pt}\selectfont $\pm$ 0.4} & 57.6 {\fontsize{6pt}{6pt}\selectfont $\pm$ 0.2}&  93.3 {\fontsize{6pt}{6pt}\selectfont $\pm$ 0.3} &  63.8 {\fontsize{6pt}{6pt}\selectfont $\pm$ 1.0} &  70.1 {\fontsize{6pt}{6pt}\selectfont $\pm$ 3.6} &  73.1 {\fontsize{6pt}{6pt}\selectfont $\pm$ 0.8} &  37.6 {\fontsize{6pt}{6pt}\selectfont $\pm$ 0.3} \\
ATST Base\cite{Li2022ATST}$^{\dagger}$ &  \textbf{\underline{92.9} {\fontsize{6pt}{6pt}\selectfont $\pm$ 0.3}} &  84.1 & 95.1 & 72.0 &  \underline{97.4} {\fontsize{6pt}{6pt}\selectfont $\pm$ 0.2} &  \underline{68.6} {\fontsize{6pt}{6pt}\selectfont $\pm$ 0.2} &  \underline{76.4} {\fontsize{6pt}{6pt}\selectfont $\pm$ 1.8} &  75.6 &  \underline{37.7} {\fontsize{6pt}{6pt}\selectfont $\pm$ 0.2} \\
\midrule
MSM-MAE\cite{niizumi2022msm-mae}  &  88.6 {\fontsize{6pt}{6pt}\selectfont $\pm$ 1.5} &  86.3 {\fontsize{6pt}{6pt}\selectfont $\pm$ 0.3} &  94.5 {\fontsize{6pt}{6pt}\selectfont $\pm$ 0.2} &  72.2 {\fontsize{6pt}{6pt}\selectfont $\pm$ 0.2} &  97.5 {\fontsize{6pt}{6pt}\selectfont $\pm$ 0.1} &  70.2 {\fontsize{6pt}{6pt}\selectfont $\pm$ 1.1} &  78.4 {\fontsize{6pt}{6pt}\selectfont $\pm$ 2.8} &  75.9 {\fontsize{6pt}{6pt}\selectfont $\pm$ 0.2} &\textbf{42.5 {\fontsize{6pt}{6pt}\selectfont $\pm$ 0.7}}\\
\addlinespace[-0.02cm] \hdashline \addlinespace[0.05cm]

M2D \small{ratio=0.6} &  89.7 {\fontsize{6pt}{6pt}\selectfont $\pm$ 0.2} &\textbf{87.6 {\fontsize{6pt}{6pt}\selectfont $\pm$ 0.2}}& {95.4 {\fontsize{6pt}{6pt}\selectfont $\pm$ 0.1}}&\textbf{73.1 {\fontsize{6pt}{6pt}\selectfont $\pm$ 0.1}}& {97.9 {\fontsize{6pt}{6pt}\selectfont $\pm$ 0.1}}&\textbf{71.7 {\fontsize{6pt}{6pt}\selectfont $\pm$ 0.3}}&  83.3 {\fontsize{6pt}{6pt}\selectfont $\pm$ 1.0} &  75.3 {\fontsize{6pt}{6pt}\selectfont $\pm$ 0.1} &  41.0 {\fontsize{6pt}{6pt}\selectfont $\pm$ 0.2}\\
M2D \small{ratio=0.7} & {89.8 {\fontsize{6pt}{6pt}\selectfont $\pm$ 0.3}}&  87.1 {\fontsize{6pt}{6pt}\selectfont $\pm$ 0.3} &  94.5 {\fontsize{6pt}{6pt}\selectfont $\pm$ 0.1} &  71.3 {\fontsize{6pt}{6pt}\selectfont $\pm$ 0.4} &  97.7 {\fontsize{6pt}{6pt}\selectfont $\pm$ 0.1} &  71.6 {\fontsize{6pt}{6pt}\selectfont $\pm$ 0.3} &\textbf{83.9 {\fontsize{6pt}{6pt}\selectfont $\pm$ 1.4}}& {76.9 {\fontsize{6pt}{6pt}\selectfont $\pm$ 1.3}}&  41.8 {\fontsize{6pt}{6pt}\selectfont $\pm$ 0.4}\\

\midrule
\textcolor{gray}{ConformerXL-P Non-RA\cite{zhang2021bigssl}} & \textcolor{gray}{-} & \textcolor{gray}{-} & \textcolor{gray}{\textbf{97.5}} & \textcolor{gray}{50.3} & \textcolor{gray}{\textbf{99.7}} & \textcolor{gray}{\textbf{88.2}} & \textcolor{gray}{-} & \textcolor{gray}{-} & \textcolor{gray}{-} \\
\textcolor{gray}{AST-Fusion\#5\#12\cite{niizumi2022composing}} &  \textcolor{gray}{\textbf{94.2}} &  \textcolor{gray}{85.5} &  \textcolor{gray}{80.4} &  \textcolor{gray}{24.9} &  \textcolor{gray}{87.6} &  \textcolor{gray}{60.7} &  \textcolor{gray}{{82.9}} &  \textcolor{gray}{77.6} &  \textcolor{gray}{34.6} \\
\textcolor{gray}{BYOL-S\cite{scheidwasserclow2021serab}} & \textcolor{gray}{-} & \textcolor{gray}{-} & \textcolor{gray}{-} & \textcolor{gray}{-} & \textcolor{gray}{-} & \textcolor{gray}{76.9} & \textcolor{gray}{-} & \textcolor{gray}{-} & \textcolor{gray}{-}\\
\bottomrule
\addlinespace[0.05cm]
\multicolumn{10}{l}{$^{\dagger}$
Underlined results were obtained in this study and \cite{niizumi2022byol-a} using publicly available pre-trained models.}\\
\end{tabular}
}
\vspace{-5pt}
\end{table*}

\begin{table}[htb!]
\vspace{-10pt}
\caption{Fine-tuning comparison with SOTA models.\\ Supervised learning method results are grayed out as a reference.}
\label{tab:results-sota-ft}
\centering
\resizebox{0.85\columnwidth}{!}{%
\begin{tabular}{lllll}
\toprule
 &     AS20K &     ESC-50 &     SPCV2 &       VC1\\
\vspace{-1pt} Model     &       mAP &     acc(\%) &  acc(\%) &    acc(\%)\\
\midrule
%Conformer\cite{Srivastava2022conformer} & 27.6 & 88.0 & - & - \\
DeLoRes-M\cite{Ghosh2022DeLoRes} & - & - & 96.0 & 62.0 \\
MAE-AST Patch/Frame\cite{Baade2022MAE-AST} & 30.6 & 90.0 & 98.0 & 63.3 \\
MaskSpec/MaskSpec-small\cite{chong2022maskspec} & 32.3 & 90.7 & 97.7 & - \\
SSAST 250/400\cite{gong2022ssast} & 31.0 & 88.8 & 98.2 & 66.6 \\
data2vec\cite{baevski2022data2vec} & 34.5 & - & - & - \\
%Audio-MAE (global)\cite{huang2022maskedlisten} & 36.6 & 93.6 & 98.3 & 94.1 \\
Audio-MAE (local)\cite{huang2022maskedlisten} & 37.1 & 94.1 & 98.3 & 94.8 \\
ATST Base\cite{Li2022ATST} &  \textbf{37.4} & - & 98.0 & 94.3 \\
\midrule
MSM-MAE\cite{niizumi2022msm-mae}  &  36.7 {\fontsize{6pt}{6pt}\selectfont $\pm$ 0.5} &  94.0 {\fontsize{6pt}{6pt}\selectfont $\pm$ 0.2} &  98.4 {\fontsize{6pt}{6pt}\selectfont $\pm$ 0.1} &\textbf{95.3 {\fontsize{6pt}{6pt}\selectfont $\pm$ 0.1}}\\
\addlinespace[-0.02cm] \hdashline \addlinespace[0.05cm]

M2D \small{ratio=0.6} &  36.8 {\fontsize{6pt}{6pt}\selectfont $\pm$ 0.1} &  94.7 {\fontsize{6pt}{6pt}\selectfont $\pm$ 0.3} &\textbf{98.5 {\fontsize{6pt}{6pt}\selectfont $\pm$ 0.0}}&  94.8 {\fontsize{6pt}{6pt}\selectfont $\pm$ 0.1} \\
M2D \small{ratio=0.7} &\textbf{37.4 {\fontsize{6pt}{6pt}\selectfont $\pm$ 0.1}}&\textbf{95.0 {\fontsize{6pt}{6pt}\selectfont $\pm$ 0.2}}&  98.5 {\fontsize{6pt}{6pt}\selectfont $\pm$ 0.1} &  94.4 {\fontsize{6pt}{6pt}\selectfont $\pm$ 1.3} \\

\midrule
\textcolor{gray}{AST (Single), AST-P/S\cite{gong2021ast}}   & \textcolor{gray}{34.7} & \textcolor{gray}{95.6} & \textcolor{gray}{98.11} & \textcolor{gray}{-} \\
\textcolor{gray}{EAT-S/M\cite{gazneli2022EAT}}   & \textcolor{gray}{-} & \textcolor{gray}{96.3} & \textcolor{gray}{98.15} & \textcolor{gray}{-} \\
\textcolor{gray}{PaSST\cite{Koutini2022passt}} & \textcolor{gray}{-} & \textcolor{gray}{96.8} & \textcolor{gray}{-} & \textcolor{gray}{-}\\
\textcolor{gray}{HTS-AT\cite{Chen2022HTS-AT}} & \textcolor{gray}{-} & \textcolor{gray}{\textbf{97.0}} & \textcolor{gray}{98.0} & \textcolor{gray}{-}\\
\bottomrule\\
\end{tabular}
}
\vspace{-15pt}
\end{table}

%%%%%%%%%%%%%%%%%
%\subsection{Experiment: Masking ratio ablations} \label{exp:mask-ratio}
\subsection{Masking Ratio Ablations} \label{exp:mask-ratio}
\begin{figure}[htbp]
  \centering
  \includegraphics[width=0.90\columnwidth]{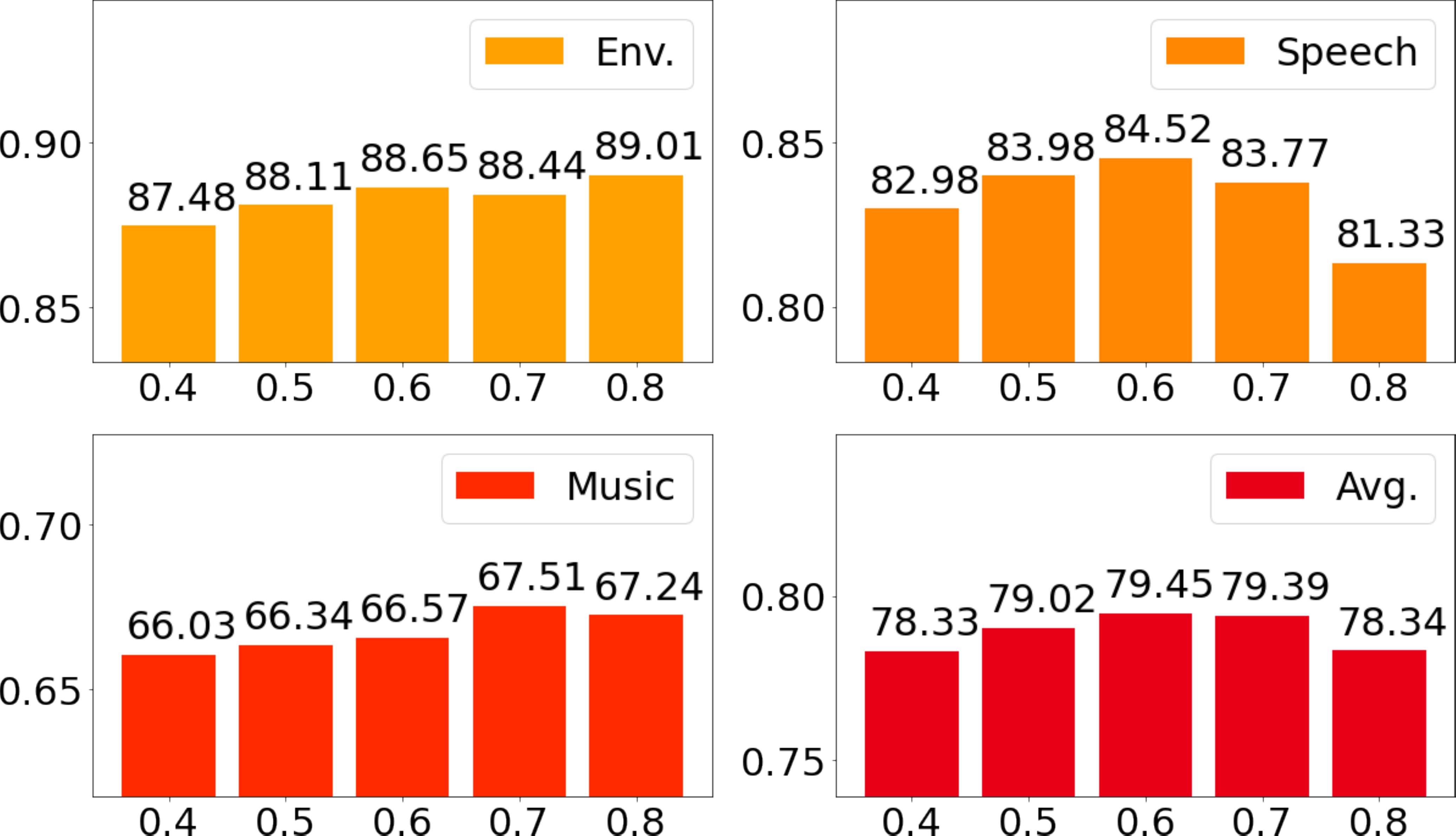} 
  \caption{Masking ratio ablations: linear evaluation results (\%).}
  \label{fig:mask-ratio-results}
  \vspace{-10pt}
\end{figure}

%%%%%%%%%%%%%%%%%
%様々なマスク率でのLinear evaluation性能を評価する。
We evaluate the impact of various masking ratios using linear evaluation.
%図4に3つのタスクグループの結果と全体の平均を示す。平均結果はマスク率0.6が最も良いことを示している。
Fig. \ref{fig:mask-ratio-results} shows the results for the three task groups and the overall average, showing the best average result (Avg.) at 0.6.
%しかしながら、3つのグループはそれぞれ違った傾向を示しており、最適なマスク率がタスクによって異なることを示している。環境音タスクでは0.8で最も良い結果となり、音声タスクでは0.6がピークである。これらに対して音楽タスクでは0.8に最大値を持つ。このように、一つの共通な最適値を設定するのが困難である。
However, the three groups show different trends, indicating that the optimal masking ratio depends on the task. The environmental sound tasks show the best result at 0.8, while the speech tasks have a peak of 0.6. The music task shows the best result at 0.7. Thus, it is difficult to set a single common optimal value.

%我々は最適なマスク率の違いは、MAEで議論されたように目的とする情報の密度の違いによるものと考えられる。音声タスクの音は連続する音素から成り、多くのマスクは予測が困難になる一方、環境音や音楽タスクは連続する長い音(エアコンの音や長い音符の楽器音など)も含まれるためマスク率が高くても予測がしやすいと考えられる。
We suspect that these optimal masking ratios are due to differences in the density of the target information, as discussed in the MAE paper.
The sounds in the speech tasks consist of successive phonemes, making many masks difficult to predict, while the environmental and music tasks include long continuous sounds (e.g., the sound of an air conditioner or instrumental sounds with long notes), making them easier to predict even at higher masking ratios.
%実験結果は学習される表現の持つ様々な音に対する有用性がマスク率により変化する、更には制御できることを示していると考えられる。
The results may indicate that the usefulness of the learned representations for various sounds can be varied and even controlled by masking ratios.

%%%%%%%%%%%%%%%%%
\subsection{Comparison with SOTA} \label{exp:sota}
%%%%%%%%%%%%%%%%%
We compare the M2D with SOTA methods in this section.
%Tables \ref{tab:results-sota-le} and \ref{tab:results-sota-ft} show the results.
%表Xに示されるLinear evaluationの結果は、我々の手法がUrbanSound8K、VoxCeleb1、CREMA-D、GTZANの4タスクで既存SSL手法に勝ることを示している。中でもUrbanSound8Kで86.8%、VoxCeleb1で72.4%となり、すべての手法に勝る結果となった。
Linear evaluation results shown in Table \ref{tab:results-sota-le} confirm that our method outperforms existing SSL methods in four tasks: UrbanSound8K, VoxCeleb1, CREMA-D, and GTZAN. Among them, M2D outperforms all previous methods with 87.6\% on UrbanSound8K, 73.1\% on VoxCeleb1, and 83.9\% on GTZAN.
%Whereas, our previous study MSM-MAE\cite{niizumi2022msm-mae} shows the best Surge result.

%表Xに示されるFine-tuningの結果は、我々の手法がESC-50、Speech commands V2で既存SSL手法に勝ることを示している。中でもSpeech commands V2は98.4%を記録し、教師あり学習を含むすべての既存手法に勝る結果となった。
The fine-tuning results in Table \ref{tab:results-sota-ft} show that our method outperforms existing SSL methods on ESC-50 and Speech commands V2 tasks and gives the same SOTA result as ATST\cite{Li2022ATST} on AudioSet20K. In particular, the result of 98.5\% on Speech commands V2 outperforms all previous methods, including supervised learning.
MSM-MAE\cite{niizumi2022msm-mae} shows a new SOTA VoxCeleb1 result, and M2D with a masking ratio of 0.6 shows a comparable result.

%M2D多くのタスクでSOTA性能を示した一方、教師あり学習に対しては環境音タスクで、音声に特化したモデルに対しては音声タスクの性能には及んでいない。これらのタスク性能向上は今後の研究課題である。
While the M2D showed SOTA performance on many tasks, it underperforms supervised learning methods (EAT\cite{gazneli2022EAT}, PaSST\cite{Koutini2022passt}, and HTS-AT\cite{Chen2022HTS-AT}) on ESC-50 and speech models (Wav2Vec2\cite{baevski2020wav2vec2}, BigSSL\cite{zhang2021bigssl}, and BYOL-S\cite{scheidwasserclow2021serab}) on the speech tasks, suggesting a future research direction.
In summary, the experiments demonstrate the effectiveness of M2D among SOTA methods.

%一方で我々の手法は、BYOLフレームワークでViTを学習するATSTにはESC-50/AS20Kの環境音認識タスクで劣り、音声タスクではWav2Vec2やBigSSLに及ばない結果がある。非-masked prediction学習や音声タスクへの強化が今後の方向性として考えられる。
%On the other hand, our method is underperforming ATST, which pre-trains ViT using the BYOL framework, in the environmental sound recognition task on ESC-50/AS20K, and is inferior to Wav2Vec2 and BigSSL in the speech tasks. Future directions may include enhancement to non-masked prediction learning and speech tasks.

%%%%%%%%%%%%%%%%%%%%%%%%%%%%%%%%%%%%%%%%%%%%%%%%%%%%
\section{Conclusion}
\label{sec:conclusion}
%%%%%%%%%%%%%%%%%%%%%%%%%%%%%%%%%%%%%%%%%%%%%%%%%%%%

In this study, we proposed a new method, Masked Modeling Duo (M2D), to learn representations directly by predicting masked patch representations using two networks. Unlike previous methods, we encode training signal representations using only masked patches rather than all input patches.
We made the M2D so that it encourages both online and target networks to model the entire input in visible and masked patch representations, respectively, achieving an effective representation.
To evaluate our method, we applied it to masked spectrogram modeling to learn general-purpose audio representations with an audio spectrogram as input. We evaluated the performance of our method on a variety of downstream tasks.
Experiments validated the effectiveness of our method, and M2D showed state-of-the-art results on UrbanSound8K, VoxCeleb1, CREMA-D, and GTZAN in linear evaluation, and on AudioSet20K, ESC-50, and Speech commands V2 in fine-tuning.
Our code is available online.

\vfill

\bibliographystyle{IEEEbib}
%\scriptsize{\bibliography{refs}}
\footnotesize{\bibliography{refs}}

\begin{thebibliography}{10}

\bibitem{he2022masked}
K.~He, X.~Chen, S.~Xie, Y.~Li, P.~Doll^^c3^^a1r, and R.~Girshick,
\newblock ``Masked autoencoders are scalable vision learners,''
\newblock in {\em CVPR}, 2022.

\bibitem{tao2022MIM:SIM}
C.~Tao, X.~Zhu, G.~Huang, Y.~Qiao, X.~Wang, and J.~Dai,
\newblock ``Siamese image modeling for self-supervised vision representation
  learning,''
\newblock {\em arXiv preprint arXiv:2206.01204}, 2022.

\bibitem{assran2022MIM:MSN}
M.~Assran, M.~Caron, I.~Misra, P.~Bojanowski, F.~Bordes, P.~Vincent, A.~Joulin,
  M.~Rabbat, and N.~Ballas,
\newblock ``Masked siamese networks for label-efficient learning,''
\newblock in {\em ECCV}, 2022.

\bibitem{chen2022MIM:CAE}
X.~Chen, M.~Ding, X.~Wang, Y.~Xin, S.~Mo, Y.~Wang, S.~Han, P.~Luo, G.~Zeng, and
  J.~Wang,
\newblock ``Context autoencoder for self-supervised representation learning,''
\newblock {\em arXiv preprint arXiv:2202.03026}, 2022.

\bibitem{elnouby2021MIM:SplitMask}
A.~El-Nouby, G.~Izacard, H.~Touvron, I.~Laptev, H.~Jegou, and E.~Grave,
\newblock ``Are large-scale datasets necessary for self-supervised
  pre-training?,''
\newblock {\em arXiv preprint arXiv:2112.10740}, 2021.

\bibitem{huang2022maskedlisten}
P.-Y. Huang, H.~Xu, J.~Li, A.~Baevski, M.~Auli, W.~Galuba, F.~Metze, and
  C.~Feichtenhofer,
\newblock ``Masked autoencoders that listen,''
\newblock in {\em NeurIPS}, 2022.

\bibitem{Baade2022MAE-AST}
A.~Baade, P.~Peng, and D.~Harwath,
\newblock ``{MAE-AST: Masked Autoencoding Audio Spectrogram Transformer},''
\newblock in {\em Interspeech}, 2022, pp. 2438--2442.

\bibitem{chong2022maskspec}
D.~Chong, H.~Wang, P.~Zhou, and Q.~Zeng,
\newblock ``Masked spectrogram prediction for self-supervised audio
  pre-training,''
\newblock {\em arXiv preprint arXiv:2204.12768}, 2022.

\bibitem{niizumi2022msm-mae}
D.~Niizumi, D.~Takeuchi, Y.~Ohishi, N.~Harada, and K.~Kashino,
\newblock ``{Masked Spectrogram Modeling using Masked Autoencoders for Learning
  General-purpose Audio Representation},''
\newblock in {\em HEAR: Holistic Evaluation of Audio Representations (NeurIPS
  2021 Competition)}, 2022, vol. 166, pp. 1--24.

\bibitem{baevski2022data2vec}
A.~Baevski, W.-N. Hsu, Q.~Xu, A.~Babu, J.~Gu, and M.~Auli,
\newblock ``data2vec: A general framework for self-supervised learning in
  speech, vision and language,''
\newblock in {\em ICML}, 2022, pp. 1298--1312.

\bibitem{Xinlei2021SimSham}
X.~Chen and K.~He,
\newblock ``Exploring simple siamese representation learning,''
\newblock in {\em CVPR}, Jun 2021.

\bibitem{grill2020byol}
J.-B. Grill, F.~Strub, F.~Altch^^c3^^a9, C.~Tallec, P.~H. Richemond,
  E.~Buchatskaya, C.~Doersch, B.~A. Pires, Z.~D. Guo, M.~G. Azar, B.~Piot,
  K.~Kavukcuoglu, R.~Munos, and M.~Valko,
\newblock ``Bootstrap your own latent - a new approach to self-supervised
  learning,''
\newblock in {\em NeurIPS}, 2020.

\bibitem{Liu2020Mockingjay}
A.~T. Liu, S.-w. Yang, P.-H. Chi, P.-c. Hsu, and H.-y. Lee,
\newblock ``Mockingjay: Unsupervised speech representation learning with deep
  bidirectional transformer encoders,''
\newblock in {\em ICASSP}, 2020, pp. 6419--6423.

\bibitem{baevski2020wav2vec2}
A.~Baevski, Y.~Zhou, A.~Mohamed, and M.~Auli,
\newblock ``wav2vec 2.0: {A} framework for self-supervised learning of speech
  representations,''
\newblock in {\em NeurIPS}, 2020.

\bibitem{Hsu2021HuBERT}
W.-N. Hsu, B.~Bolte, Y.-H.~H. Tsai, K.~Lakhotia, R.~Salakhutdinov, and
  A.~Mohamed,
\newblock ``{HuBERT}: Self-supervised speech representation learning by masked
  prediction of hidden units,''
\newblock {\em IEEE/ACM Trans. Audio, Speech, Language Process.}, p.
  3451^^e2^^80^^933460, 2021.

\bibitem{zhang2021bigssl}
Y.~Zhang, D.~S. Park, W.~Han, J.~Qin, A.~Gulati, J.~Shor, A.~Jansen, Y.~Xu,
  Y.~Huang, S.~Wang, et~al.,
\newblock ``{BigSSL}: Exploring the frontier of large-scale semi-supervised
  learning for automatic speech recognition,''
\newblock {\em IEEE J. Sel. Top. Signal Process.}, vol. 16, no. 6, pp.
  1519^^e2^^80^^931532, 2022.

\bibitem{gong2022ssast}
Y.~Gong, C.-I. Lai, Y.-A. Chung, and J.~Glass,
\newblock ``{SSAST}: Self-supervised audio spectrogram transformer,''
\newblock in {\em AAAI}, 2022, vol.~36, pp. 10699--10709.

\bibitem{wang2022universal}
L.~Wang, P.~Luc, Y.~Wu, A.~Recasens, L.~Smaira, A.~Brock, A.~Jaegle, J.-B.
  Alayrac, S.~Dieleman, J.~Carreira, and A.~van~den Oord,
\newblock ``Towards learning universal audio representations,''
\newblock in {\em ICASSP}, 2022, pp. 4593--4597.

\bibitem{Ghosh2022DeLoRes}
S.~Ghosh, A.~Seth, and S.~Umesh,
\newblock ``Decorrelating feature spaces for learning general-purpose audio
  representations,''
\newblock {\em IEEE J. Sel. Topics Signal Process.}, vol. 16, no. 6, pp.
  1402--1414, 2022.

\bibitem{niizumi2022byol-a}
D.~Niizumi, D.~Takeuchi, Y.~Ohishi, N.~Harada, and K.~Kashino,
\newblock ``{BYOL for Audio}: Exploring pre-trained general-purpose audio
  representations,''
\newblock {\em IEEE/ACM Trans. Audio, Speech, Language Process.}, vol. 31, pp.
  137^^e2^^80^^93151, 2023.

\bibitem{scheidwasserclow2021serab}
N.~Scheidwasser-Clow, M.~Kegler, P.~Beckmann, and M.~Cernak,
\newblock ``{SERAB}: A multi-lingual benchmark for speech emotion
  recognition,''
\newblock in {\em ICASSP}, 2022, pp. 7697--7701.

\bibitem{Li2022ATST}
X.~LI and X.~Li,
\newblock ``{ATST: Audio Representation Learning with Teacher-Student
  Transformer},''
\newblock in {\em Interspeech}, 2022, pp. 4172--4176.

\bibitem{gong2021ast}
Y.~Gong, Y.-A. Chung, and J.~Glass,
\newblock ``{AST}: Audio spectrogram transformer,''
\newblock in {\em Interspeech}, 2021, pp. 571--575.

\bibitem{gazneli2022EAT}
A.~Gazneli, G.~Zimerman, T.~Ridnik, G.~Sharir, and A.~Noy,
\newblock ``End-to-end audio strikes back: Boosting augmentations towards an
  efficient audio classification network,''
\newblock {\em arXiv preprint arXiv:2204.11479}, 2022.

\bibitem{Koutini2022passt}
K.~Koutini, J.~Schl^^c3^^bcter, H.~Eghbal-zadeh, and G.~Widmer,
\newblock ``Efficient training of audio transformers with patchout,''
\newblock in {\em Interspeech}, 2022, pp. 2753--2757.

\bibitem{Chen2022HTS-AT}
K.~Chen, X.~Du, B.~Zhu, Z.~Ma, T.~Berg-Kirkpatrick, and S.~Dubnov,
\newblock ``{HTS-AT}: A hierarchical token-semantic audio transformer for sound
  classification and detection,''
\newblock in {\em ICASSP}, 2022, pp. 646--650.

\bibitem{ViT}
A.~Dosovitskiy, L.~Beyer, A.~Kolesnikov, D.~Weissenborn, X.~Zhai,
  T.~Unterthiner, M.~Dehghani, M.~Minderer, G.~Heigold, S.~Gelly, J.~Uszkoreit,
  and N.~Houlsby,
\newblock ``An image is worth 16x16 words: Transformers for image recognition
  at scale,''
\newblock in {\em ICLR}, 2021.

\bibitem{gemmeke2017audioset}
J.~F. Gemmeke, D.~P.~W. Ellis, D.~Freedman, A.~Jansen, W.~Lawrence, R.~C.
  Moore, M.~Plakal, and M.~Ritter,
\newblock ``{Audio Set}: An ontology and human-labeled dataset for audio
  events,''
\newblock in {\em ICASSP}, 2017, pp. 776--780.

\bibitem{piczak2015esc50}
K.~J. Piczak,
\newblock ``{ESC}: {Dataset} for {Environmental Sound Classification},''
\newblock in {\em ACM-MM}, 2015, pp. 1015--1018.

\bibitem{salamon2014urbansound}
J.~Salamon, C.~Jacoby, and J.~P. Bello,
\newblock ``A dataset and taxonomy for urban sound research,''
\newblock in {\em ACM-MM}, 2014, pp. 1041--1044.

\bibitem{speechcommandsv2}
P.~{Warden},
\newblock ``{Speech Commands: A Dataset for Limited-Vocabulary Speech
  Recognition},''
\newblock {\em arXiv preprint arXiv::1804.03209}, Apr. 2018.

\bibitem{voxceleb}
A.~Nagrani, J.~S. Chung, and A.~Zisserman,
\newblock ``Voxceleb: A large-scale speaker identification dataset,''
\newblock in {\em Interspeech}, 2017, pp. 2616--2620.

\bibitem{voxforge}
K.~MacLean,
\newblock {\em “Voxforge”}, 2018,
\newblock Available at \url{http://www.voxforge.org/home}.

\bibitem{cao2014cremad}
H.~Cao, D.~G. Cooper, M.~K. Keutmann, R.~C. Gur, A.~Nenkova, and R.~Verma,
\newblock ``{CREMA-D}: Crowd-sourced emotional multimodal actors dataset,''
\newblock {\em IEEE Trans. Affective Comput.}, vol. 5, no. 4, 2014.

\bibitem{gt2013gtzan}
G.~Tzanetakis and P.~Cook,
\newblock ``Musical genre classification of audio signals,''
\newblock {\em IEEE Speech Audio Process.}, vol. 10, no. 5, 2002.

\bibitem{nsynth2017}
J.~Engel, C.~Resnick, A.~Roberts, S.~Dieleman, M.~Norouzi, D.~Eck, and
  K.~Simonyan,
\newblock ``Neural audio synthesis of musical notes with {W}ave{N}et
  autoencoders,''
\newblock in {\em ICML}, 2017.

\bibitem{turian2021torchsynth}
J.~Turian, J.~Shier, G.~Tzanetakis, K.~McNally, and M.~Henry,
\newblock ``One billion audio sounds from {GPU}-enabled modular synthesis,''
\newblock in {\em DAFx2020}, 2021.

\bibitem{loshchilov2016sgdr}
I.~Loshchilov and F.~Hutter,
\newblock ``{SGDR}: Stochastic gradient descent with warm restarts,''
\newblock in {\em ICLR}, 2017.

\bibitem{zhang2018mixup}
H.~Zhang, M.~Cisse, Y.~N. Dauphin, and D.~Lopez-Paz,
\newblock ``mixup: Beyond empirical risk minimization,''
\newblock in {\em ICLR}, 2018.

\bibitem{niizumi2022composing}
D.~Niizumi, D.~Takeuchi, Y.~Ohishi, N.~Harada, and K.~Kashino,
\newblock ``Composing general audio representation by fusing multilayer
  features of a pre-trained model,''
\newblock in {\em EUSIPCO}, 2022, pp. 200--204.

\bibitem{ImageNet}
J.~Deng, W.~Dong, R.~Socher, L.-J. Li, K.~Li, and L.~Fei-Fei,
\newblock ``Imagenet: A large-scale hierarchical image database,''
\newblock in {\em CVPR}, 2009.

\end{thebibliography}
%\bibliography{refs}

\newpage

\appendix

\noindent \textbf{NOTE: The following appendix will not be on the ICASSP paper. Please cite the arXiv version if you reference the appendix of this paper.}
\vspace{0.5cm}

%%%%%%%%%%%%%%%%%%%%%%%%%%%%%%%%%%%%%%%%%%%%%%%%%%%%
\section{Experiments with Images}
%%%%%%%%%%%%%%%%%%%%%%%%%%%%%%%%%%%%%%%%%%%%%%%%%%%%

%画像に対するM2Dの有効性を検証するため、ImageNetを使った標準的な性能評価を実施する。
%%ImageNet-1Kを使って事前学習を行い、その後fine-tuningを行った。我々はMAE同様に画像サイズ224x224のtop-1 accuracyを報告する。
We validate the effectiveness of our M2D for images by conducting evaluations on ImageNet\cite{ImageNet}.
We pre-trained on ImageNet-1K\cite{ImageNet}, followed by fine-tuning. We report top-1 validation accuracy for a single crop image of $224 \times 224$ same as in the MAE\cite{he2022masked}.

%この実験においても4章同様にMAEのコードをベースとして最小限の変更にとどめることで、実験対象のみの差分の比較を可能にした。バックボーンにも4章同様ViT-Baseを利用した。
%事前学習のパラメータはエポック数300、バッチサイズ2048、とした他はマスク率0.75を含めMAEと同じ設定とした。ターゲットネットワーク更新のEMAパラメータTAUも音響信号での実験同様、学習開始時0.99995から終了時0.99999まで線形補間した。
%fine-tuningはMAEのソースコードをそのまま利用しパラメータも同一に設定した。
We used the MAE code as a base, as in Section \ref{sec:experiments} using audio, and made minimal changes to it, allowing comparison of differences only in the experimental subjects of interest. We also used ViT-Base\cite{ViT} for the backbone.
For pre-training, we set the number of epochs as 300 and the batch size as 2048. All other settings were the same as in the MAE, including the masking ratio of 0.75. The EMA decay rate $\tau$  for the target network update was also the same as in Section \ref{sec:experiments}, linearly interpolated from 0.99995 at the start of training to 0.99999 at the end.
For fine-tuning, we used the same source code and parameters as in the MAE.

%Table 6は、MAE、M2D、M2Dのターゲットエンコーダに全てのパッチを入力した場合、3モデルの結果を示す。従来法であるターゲットに全てのパッチを入力するM2D variantではMAE同等の性能であり、我々のM2Dはこれらの性能を上回ることを確認した。これらの結果は、提案手法が画像においても有効であることを検証するものである。
We compare three models, MAE, M2D, and an M2D variant, that feed all patches to the target encoder. Table \ref{tab:results-imagenet} shows the results. The results show the performance of the M2D variant with all patches input to the target, which is conventional, is comparable to MAE, and our M2D outperforms these methods, validating that our method is also effective on images in addition to audios.

\begin{table}[tbh!]
%\vspace{-10pt}
\caption{Fine-tuning results on ImageNet-1K.}
\label{tab:results-imagenet}
\centering
\resizebox{\columnwidth}{!}{%
\begin{tabular}{llll}
\toprule
Model & Target & Target & Top-1 \\
\vspace{-1pt}      & encoder &  input   &    acc(\%) \\
\midrule
MAE &  & - & 83.22 {\fontsize{6pt}{6pt}\selectfont $\pm$ 0.024}\\
%\addlinespace[-0.02cm] \hdashline \addlinespace[0.05cm]
M2D variant (conventional) & \checkmark  & All patches & 83.22 {\fontsize{6pt}{6pt}\selectfont $\pm$ 0.085}\\
M2D (ours) & \checkmark & Masked patches only & 83.35 {\fontsize{6pt}{6pt}\selectfont $\pm$ 0.211}\\
\bottomrule\\
\end{tabular}
}
\vspace{-15pt}
\end{table}

\end{document}